\documentclass[preprint]{elsarticle} 

\usepackage{lineno,hyperref}

\modulolinenumbers[5] 

\usepackage{arydshln} 
\usepackage{amsmath,amsthm,verbatim,amssymb,amsfonts,amscd, graphicx}
\def\be{\begin{equation}}
\def\ee{\end{equation}}
\def\ba{\begin{eqnarray}}
\def\ea{\end{eqnarray}}
\newcommand{\m}{\rm{m}}
\newcommand{\fr}[2]{\frac{#1}{#2}}
\newcommand{\DE}{\rm{DE}}

\newcommand{\jcap}{J.\ Cosmol.\ Astropart.\ Phys\,}
\newcommand{\physrep}{Phys.\ Rept \,}
\newcommand{\eff}{\rm{eff}}
\newcommand{\crt}{\rm{cr}}
\newcommand{\HS}{\rm{HS}}
\usepackage{multirow} 
\usepackage{booktabs}
\usepackage{mathtools} 
\usepackage{multicol} 

\begin{document}

\begin{frontmatter}

\title{Reconstruction of f(R) gravity models from observations} 

\author{Seokcheon Lee\fnref{myfootnote}}
\address{The Research Institute of Natural Science, Gyeongsang National University, Jinju 52828, Republic of Korea}
\fntext[myfootnote]{skylee2@gmail.com}




\begin{abstract}
We investigate the method for the reconstruction of $f(R)$ gravity models both from the background evolution observations and from the large scale structure measurements. Due to the lack of the first principles, one needs to rely on the observations to build the $f(R)$ gravity models. We show that the general $f(R)$ models can be specified if the 5\% accuracy level  large scale structure formation observations will be available in the near future. 
\end{abstract}


\end{frontmatter}


\section{Introduction}

The current accelerating expansion of the Universe can be well explained by the cosmological constant. However the cosmological constant suffers from the so-called the fine tuning and the coincidence problems.  Thus, both dark energy (DE) with an exotic equation of state (e.o.s) and a modification of gravity (MG) on large scales are investigated as the alternative solutions to the cosmological constant \cite{2012PhR...513....1C, 2014arXiv1401.0046M, 2016ARNPS..66...95J}. Two theories are distinguished by the violation of the so-called the strong equivalence principle and one is not able to probe whether this is correct or not. Thus, the origin of the accelerating expansion of the Universe can be probed by observations.   

The background evolution of both DE and MG models are degenerated, described by the Hubble parameter $H$ and the integrations of its inverse with respect to the redshift. This degeneracy can be broken by observing the evolution of cosmological perturbations obtained from the large scale structure formation. Thus, one needs to have accurate observations both in the background evolution and in the large scale structure in order to reconstruct the viable MG models. Traditionally, one specifies the specific form of MG and fits the model parameters from observations of the background evolution. From these fixed parameters of the specific models, one predicts the observations of the large scale structure. However, this method is quite ambiguous. Also, if one includes the errors on the measurements then the probing the specific form of model is quite difficult. 

The background evolution of the Universe can be specified by the DE e.o.s, $\omega$. To be used for the various models, one usually adopts the parametrization of $\omega$ \cite{2001IJMPD..10..213C, 2003PhRvL..90i1301L}. Also the evolution of the cosmological perturbation is usually parametrized by using the so-called growth rate index parameter, $\gamma$. We also adopt the specific form of $\gamma$ to generalize the growing of the matter perturbation \cite{2011JCAP...03..021L}. 

In this {\it Letter}, we focus on the $f(R)$ gravity models among various MG models. In stead of specifying the form of $f(R)$-gravity models, we reconstruct the model from the observed quantities of parametrization of $\omega$ and $\gamma$. In the next section, we briefly review both the background and the perturbation equations in general $f(R)$ gravity models in the metric formalism \cite{2010RvMP...82..451S}. We show the reconstruction method of $f(R)$-gravity models from the observed quantities and the results in Sec. 3. We derive detail formulae for the model functions as a function of $\omega$ and $\gamma$ in the appendix.          

\section{Review on Background and Perturbation Evolutions}

In this section, we review the general feature of f(R) gravity models. Both the background evolution equations and the matter perturbation evolution equations are reviewed. 

\subsection{Model}

The action of the general f(R) gravity theories are given by \cite{1980PhLB...91...99S}
\be S = \int d^4 x \sqrt{-g} \left( \fr{f(R)}{16 \pi G} + {\cal L}_{\m} \right) \equiv \int d^4 x \sqrt{-g} \left( \fr{R + \tilde{f}(R)}{16 \pi G} + {\cal L}_{\m} \right)  \, , \label{S} \ee
where $f(R)$ is a general function of the Ricci scalar R and ${\cal L}_{\m}$ is the matter Lagrangian. The Ricci scalar is given by 
\be R = 6 \left(H^2 + \dot{H} \right) = 8 \pi G \Bigl( \rho_{\m} + \left( 1 - 3 \omega_{\eff} \right) \rho_{\eff} \Bigr) \, , \label{R} \ee
where $\rho_{\m} (\rho_{\eff})$ is the mass density of the matter (the effective dark energy) component and $\omega_{\eff}$ is the e.o.s of the effective DE which is derived from the modification of the Einstein-Hilbert action. It has been considered that the form of $f(R)$ can be constrained in order to satisfy both cosmological and solar-system tests \cite{2007PhRvD..76f4004H}. It has been believed that following two conditions are required to satisfy observational tests 
\ba \lim_{R\to\infty} \tilde{f}(R) &=& \text{const.} \,\, , \nonumber \\
\lim_{R\to 0} \tilde{f}(R) &=& 0 \,\, , \label{finfty0} \ea   
which lead to several viable $f(R)$ models shown in the table \ref{tab-1}. 

\begin{table}[h]
\centering
\begin{tabular}{c ccc}
\hline \hline \\
Model		& $f(R)$ 	& parameters & Ref \\[2ex]		
\hline
\\[0.2ex]
Hu-Sawicki & $R-  R_{\text{HS}} \fr{c_1 (R/ R_{\text{HS}})^{p}}{c_2 (R/ R_{\text{HS}})^{p} +1} $ & $R_{\text{HS}} >0, c_1, c_2, p >0$ & \cite{2007PhRvD..76f4004H} \\[3ex]
\hdashline
\\[0.2ex]
Starobinsky & $R - \lambda R_{\text{S}} \left[ 1 - \left(1+ \fr{R^2}{R_{\text{S}}^2} \right)^{-n} \right] $ & $\lambda > 0, n > 0, \lambda R_{\text{S}} \sim 2 \Lambda$ & \cite{2007JETPL..86..157S} \\[3ex]
\hdashline
\\[0.2ex]
Tsujikawa & $R - \mu R_{\text{T}} \tanh \left[\fr{R}{R_{\text{T}}} \right] $ & $\mu >0, R_{\text{T}} > 0$ & \cite{2008PhRvD..77b3507T} \\[3ex]
\hdashline
\\[0.2ex]
Exponential Gravity & $R - \beta R_{\text{E}} \left( 1 - e^{-R}{R_{\text{E}}} \right)$ & $\beta >1, R_{\text{E}} > 0$ & \cite{2008PhRvD..77d6009C} \\[3ex]
\hline
\end{tabular}
\caption{Viable $f(R)$ gravity models.} 
\label{tab-1}
\end{table}
There are also constraints on the derivatives of $f(R)$ to satisfy the attractive gravity and the stability of the model.
\ba F &\equiv& f_{,R} = \fr{\partial f}{\partial R} = 1 + \tilde{f}_{,R} > 0 \, , \label{F} \\
F_{,R} &=& f_{,RR} = \fr{\partial^2 f}{\partial R^2} = \tilde{f}_{,RR} > 0 \,\, \text{for} \,\, R \gg m^2 \, . \label{FR} \ea
One can refer the recent review on these models \cite{2017arXiv171005634P}.

\subsection{Background Evolution Equations}
We are interested in the evolution of the Universe after the matter dominated epoch. The gravitational field equation is obtained from the variation of action, Eq(\ref{S}) with respect to the metric 
\be F R_{\mu\nu} -\fr{1}{2} f g_{\mu\nu} - F_{,\mu ; \nu} + \Box F g_{\mu\nu} = 8\pi G T^{(\m)}_{\mu\nu} \, , \label{Feq} \ee 
where $T^{(\m)}_{\mu\nu}$ is an energy-momentum tensor of the pressureless matter. In a flat Friedmann-Lema$\hat{i}$tre-Robertson-Walker (FLRW) metric with a scale factor a, one obtain the following background equations
\ba 3 F_{0} H^2 &=& \fr{1}{2} \left( F R - f \right) - 3 H \dot{F} + 3 H^2 \left( F_{0} - F \right) + 8 \pi G \rho_{\m} \, , \label{G00} \\
- 2 F_{0} \dot{H} &=& \ddot{F} - H \dot{F} - 2 \left( F_{0} - F \right) \dot{H} + 8 \pi G \left( \rho_{\m} + P_{\m} \right) \, ,\label{Gii} \\
\dot{\rho}_{\m} &=& - 3 H \left( \rho_{\m} + P_{\m} \right) \label{BI} \, , \ea
where $F_0$ denotes the present value of $F$ and the pressure of matter component, $P_{\m}$ will be ignored.

\subsection{Perturbation Equations}
The equation  for the matter perturbation at the sub-horizon scale limit, $\fr{k^2}{a^2 H^2} \gg 1$ is given by \cite{2002PhRvD..66h4009H}
\be \ddot{\delta}_{\m} + 2H \dot{\delta}_{\m} - \fr{4\pi G}{F} \rho_{\m} \left( \fr{1 + 4 M}{1+3M} \right) \delta_{\m} = 0 \,\, , \text{where} \,\, M = \fr{k^2}{a^2} \fr{F_{,R}}{F} \label{dotdeltam} \, . \ee
The growth rate of the matter perturbation is well parameterized as 
\be d \ln \delta_{\m} / d \ln a \equiv \Omega_{\m}^{\gamma} \, , \label{growthrate} \ee where the growth rate index, $\gamma$ can be parameterized as \cite{2011JCAP...03..021L}
\be \gamma = \gamma_{0} + \gamma_{a} \left(1 - e^{n} \right) \, . \label{gamma} \ee 

The term in the perturbation equation, $M$ can be divided for three limits

\begin{enumerate}[(I)]

\item $M \ll 1$ : \\
These limit corresponds scales where the large scale structure is formed, ($R M \ll \fr{c^2 k^2}{a^2 R} R M \ll 1$). The solutions for the matter perturbation are given by $\delta_{\m} = \exp \left[ \int \alpha dn \right]$ and using the approximation $|\alpha'| \ll \alpha^2$, one obtains the growing mode solution \cite{2007PhRvD..76b3514T}
\be \alpha_{+} = 1 + \fr{3}{5} M = 1 + \fr{3}{5} \fr{c^2 k^2}{a^2 R} \fr{R F_{,R}}{F} \equiv 1 + \fr{3}{5} \fr{c^2 k^2}{a^2 R} m \, , \label{alphap} \ee
where $m$ is the dimensionless quantity. If $m$ is constant, then $a^2 R \propto a^{-1}$ during the matter dominated era.  Thus, the growing mode of the matter perturbation is given by
\be \delta_{\m +} \propto a^{1+ \left(3/5\right) \beta_{+}} \,\, , \text{where} \,\, \beta_{+} \equiv \fr{M}{n} \, . \label{deltam} \ee
$F_{,R} = \fr{F'}{R'} > 0$ at high $R$ is required for the stability of the solution. This means $F' < 0$ at high $z$ because $R' < 0$. \\

\item $M = 0$ : \\
In this case, $F = \text{const.}$ ($F_{,R} = 0$).  This case is same as the $\Lambda$CDM model with general relativity.  \\

\item $M \gg 1$ : \\
For $M \geq 10$, one can approximate
\be \fr{1+4M}{1+3M} \approx \fr{4}{3} \label{Mapprox1} \ee
with less than 1\% error. In this case, $\omega_{\eff} = \fr{1}{3}$ and $\Omega_{\m}=2$ when $\fr{R F_{,R}}{F} = -\fr{1}{4}$ which corresponds to a $\phi$ matter dominated epoch ($\phi$MDE) \cite{2007PhRvL..98m1302A}. We will use the large scale structure formation observation to constrain the model and thus we will not consider this limit.  

\end{enumerate}

\subsection{Field Equations}
Impact of $f(R)$ can be investigated as a field equation for $F$.  The trace of Eq.(\ref{Feq}) is given by
\be \fr{\partial V_{\eff}}{\partial F} \equiv \Box F = \fr{1}{3} \left( 2f - R F - 8 \pi G \rho_{\m} \right) \, . \label{BoxF} \ee
Thus, the effective potential has an extremum at
\be 2f - RF = R + 2 \tilde{f} - R \tilde{f}_{R} = 8 \pi G \rho_{\m} \, . \label{R0} \ee


\section{Model Reconstruction from Observations}
It is convenient to rewrite the both background evolution equations (\ref{G00})-(\ref{Gii}) and the matter perturbation equation (\ref{dotdeltam}) as a function of the number of e-folding, $n \equiv \ln a$. Then both Friedmann equations and the matter perturbation equations become
\ba 3 \fr{H^2}{H_{0}^2} &=& \fr{1}{2} \left( \fr{F}{F_0} \fr{R}{H_0^2} - \fr{f}{F_0 H_0^2} \right) - 3\fr{H^2}{H_0^2} \left(\fr{F'}{F_0} + \fr{F}{F_0} - 1 - \Omega_{\m} \right)  \, , \label{G00n2} \\
 -2 \fr{H'}{H} &=& \left( \fr{F''}{F_0} - \fr{F'}{F_0} \right) + \fr{H'}{H} \left(\fr{F'}{F_0} +2\fr{F}{F_0} - 2 \right) + 3 \Omega_{\m} \, , \label{Giin2} \\
 \delta_{\m}^{''} &=& - \left( 2 + \fr{H'}{H} \right) \delta_{\m}' + \fr{3}{2} \fr{F_{0}}{F} \Omega_{\m} \left( \fr{1+4M}{1+3M} \right) \delta_{\m} \, , \label{deltamn} \ea
as shown in Eqs. (\ref{G00n2A}), (\ref{Giin2A}), and (\ref{deltamnA}).  

 Now one can rewrite the model functions as functions of observed quantities $\omega$ and $\gamma$. One can refer the appendix for the detail derivations of this section. We adopt the so-called Chevallier-Polarski-Linder (CPL) parameterization of the DE e.o.s $\omega_{\DE} = \omega_{0} + \omega_{a} \left(1 - e^{n}\right)$ \cite{2001IJMPD..10..213C, 2003PhRvL..90i1301L} to match the background evolution equations (\ref{G00n2})-(\ref{Giin2}) with (\ref{H2oH02A})-(\ref{HoHpA}). 
 \ba \fr{H^2}{H_0^2} &=& \fr{\rho_{\m}}{\rho_{\crt 0}} \left( 1 + \fr{\rho_{\DE}}{\rho_{\m}} \right) = \Omega_{\m 0} e^{-3n} \left(1+g\left[\Omega_{\m0}, \omega_{0}, \omega_{a},n \right] \right) \label{H2oH02} \, , \\
\fr{H'}{H} &=& -\fr{3}{2} \left( 1 + \omega \Omega_{\DE} \right) \equiv -\fr{3}{2} \left(1 + Q\left[\Omega_{\m0}, \omega_{0}, \omega_{a}, n \right]  \right)\label{HoHp} \, . \ea 
From these equations, one can relate model functions $f$ and $F$ with the DE e.o.s, $\omega$. However, one is not able to constrain the evolutions of model functions with these equations only. However, one also can parametrize the growth rate of the matter perturbation as $d \ln \delta_{\m} / d \ln a \equiv \Omega_{\m}^{\gamma}$ and we adopt the parametrization of the growth rate index parameter as $\gamma = \gamma_{0} + \gamma_{a} \left(1 - e^{n} \right) $ given in \cite{2011JCAP...03..021L}. One more equation required to fully constrain the model functions is obtained from the matter perturbation equation as shown in Eqs.(\ref{PA})-(\ref{FpoF0A})
\ba \fr{F}{F_0} \equiv \fr{F\left[\Omega_{\m0}, \omega_{0}, \omega_{a}, \gamma_{0}, \gamma_{a}, k, n \right] }{F\left[\Omega_{\m0}, \omega_{0}, \omega_{a}, \gamma_{0}, \gamma_{a}, k, 0 \right]} &=& \fr{3}{2} \fr{\Omega_{\m}}{{\cal P}} \left(\fr{1+4M}{1+3M}\right) \label{FoF0n} \, , \\ 
{\cal P}\left[\Omega_{\m0}, \omega_{0}, \omega_{a}, \gamma_{0}, \gamma_{a}, n \right]  &\equiv& \Omega_{\m}^{\gamma} \left( \Omega_{\m}^{\gamma} + \gamma' \ln \Omega_{\m} + 3 \gamma Q + \fr{\left(1-3Q\right)}{2} \right) \label{P} \, , \\
M = \fr{k^2}{a^2 H_{0}^2} \fr{H_0^2}{R'} \fr{F'}{F} &=& \fr{1-A}{3A -4} \,\,\,\, \text{where} \,\, A = \fr{2}{3}\fr{F}{F_{0}} \fr{{\cal P}}{\Omega_{\m}} \label{M} \, , \\
\fr{F'}{F_{0}} = \fr{a^2 H_{0}^2}{k^2} \fr{R'}{H_{0}^2} M \fr{F}{F_0}  &=& \fr{a^2 H_{0}^2}{k^2} \fr{R'}{H_{0}^2} \left( \fr{1-A}{3A-4} \right) \fr{F}{F_0} \label{FpoF0} \, . \ea 
All of the quantities in the above equations $F/F_{0}$, ${\cal P}$, $M$, and $A$ are dimensionless. 
 
We list the reconstructed model functions as a function of $\omega_0, \omega_{a}, \gamma_0$, and $\gamma_{a}$ in the table \ref{tab-4} in the appendix. Observational measurements can be parameterized by these values. However, the measurement on $\gamma_{0}$ and $\gamma_{a}$ from galaxy redshift surveys are still not accurate enough. Thus, we first assume that one can measure the exact values of ($\gamma_{0}$, $\gamma_{a}$) when ($\omega_{0}$, $\omega_{a}$) measured very accurately. We numerically calculate the ($\gamma_{0}$, $\gamma_{a}$) values from Eq.(\ref{deltamn}) for the given values of $\Omega_{\m0}$, $\omega_{0}$, $\omega_{a}$, and $M_{0}$. We assume the $\Lambda$CDM like background evolution ({\it i.e.} ($\omega_{0}, \omega_{a}) = (-1, 0)$) for the different values of $\Omega_{\m0}$ to obtain the values of ($\gamma_{0}$, $\gamma_{a}$). One also need to specify the present value of $M_{0}$ to obtain the growth index parameters ($\gamma_{0}$, $\gamma_{a}$). The best fit values for the different models are given in the table \ref{tab-2}. For the given value of the matter energy density contrast, $\Omega_{\m0}$ the magnitudes of $\gamma_{0}$ are increased as the magnitudes of $M_{0}$ decrease. Also as the magnitudes of $M_0$ decrease, so do the absolute values of $\gamma_{a}$. This is due to the fact that model gets close to $\Lambda$CDM as $M$ decreases and thus the growth index parameter of $f(R)$ model becomes that of $\Lambda$CDM. For example, ($\gamma_{0}, \gamma_{a}$) = ($0.519, -0.162$) when $\Omega_{\m0} = 0.32$ and $M_0 = 0.1$. But they become ($0.554, -0.016$) for the same value of $\Omega_{\m0}$ when $M_0 = 10^{-4}$ as shown in the table \ref{tab-2}. One can also find that the values of $\gamma_{0}$ is increased as those of $\Omega_{\m0}$ decrease for the same value of $M_0$. With the given values of  $\omega_0, \omega_a, \gamma_{0}$, and $\gamma_{a}$ one can solve the differential equation (\ref{FpoF0}) with the initial condition $\fr{F}{F_{0}} = 1$ to reconstruct the $f(R)$ models. 

\begin{table}
\centering
  \begin{tabular}{c cc cc cc}
    \hline \hline \\
    \multirow{2}{*}{$M_0$} &
      \multicolumn{2}{c}{$\Omega_{\m0} = 0.32$}  & \multicolumn{2}{c}{$\Omega_{\m0} = 0.30$} & \multicolumn{2}{c}{$\Omega_{\m0} = 0.27$} \\[1ex]
    & $\gamma_{0}$ & $\gamma_{a}$ & $\gamma_{0}$ & $\gamma_{a}$ & $\gamma_{0}$ & $\gamma_{a}$ \\[1ex]
    \hline
    \\[0.1ex]
    $10^{-1}$ & $0.519$ & $-0.162$ & $0.522$ & $-0.151$ & $0.525$  & $-0.135$ \\[1ex]
    \hdashline
    \\[0.1ex]
     $5 \times 10^{-2}$  & $0.534$ & $-0.099$ & $0.536$ & $-0.093$ & $0.538$  & $-0.085$ \\[1ex]
      \hdashline
    \\[0.1ex]
     $10^{-2}$  & $0.550$ & $-0.035$ & $0.550$ & $-0.034$ & $0.552$  & $-0.033$ \\[1ex]
      \hdashline
    \\[0.1ex]
     $10^{-3}$  & $0.554$ & $-0.017$ & $0.554$ & $-0.018$ & $0.555$  & $-0.019$ \\[1ex]
      \hdashline
    \\[0.1ex]
     $10^{-4}$  & $0.554$ & $-0.016$ & $0.555$ & $-0.016$ & $0.555$  & $-0.017$ \\[1ex]
    \hline
    \end{tabular}
\caption{The numerically obtained values of the growth index parameters for the different values of $M_0$ and $\Omega_{\m0}$ when ($\omega_{0}, \omega_{a}$) = (-1, 0).} 
\label{tab-2} 
\end{table}

We repeat the same numerical calculation to obtain the best fit values of ($\gamma_{0}, \gamma_{a}$) for the same models. However, we consider the measurement errors on the growth rate, $\fr{d \ln \delta}{d \ln a}$ at this moment. We assume the measurement error on the growth rate from -10 \% to + 10 \% when we obtain the growth index parameters ($\gamma_{0}, \gamma_{a}$). These are shown in the table \ref{tab-3}. The values of ($\gamma_{0}, \gamma_{a}$) are increased when the growth index values are decreased by including the decreased values of the growth index due to the measurement errors. For example, the measured values of ($\gamma_0, \gamma_a$) should be (0.600, 0.186) for $\Omega_{\m0} =0.32$ and $M_0 = 0.1$ model when the measured value of $\fr{d \ln \delta}{d \ln a}$ is 10 \% smaller than true value. When the measurement on the growth index is larger than the true value of it, then the values of ($\gamma_{0}, \gamma_{a}$) are smaller than the true values of the growth index parameters. 

\begin{table}
\centering
  \begin{tabular}{c cc cc cc}
    \hline \hline \\
    \multirow{2}{*}{Measurement error} &
      \multicolumn{2}{c}{$\Omega_{\m0} = 0.32$}  & \multicolumn{2}{c}{$\Omega_{\m0} = 0.30$} & \multicolumn{2}{c}{$\Omega_{\m0} = 0.27$} \\[1ex]
   in $\fr{d \ln \delta}{d \ln a} \, (\%)$ & $\gamma_{0}$ & $\gamma_{a}$ & $\gamma_{0}$ & $\gamma_{a}$ & $\gamma_{0}$ & $\gamma_{a}$ \\[1ex]
    \hline
    \\[0.1ex]
    $-10$ & $0.600$ & $0.186$ & $0.598$ & $0.169$ & $0.596$  & $0.146$ \\[1ex]
    \hdashline
    \\[0.1ex]
     $-5$  & $0.559$ & $0.008$ & $0.559$ & $0.005$ & $0.559$  & $0.001$ \\[1ex]
      \hdashline
    \\[0.1ex]
     Exact  & $0.519$ & $-0.162$ & $0.522$ & $-0.151$ & $0.525$  & $-0.135$  \\[1ex]
      \hdashline
    \\[0.1ex]
     $+5$  & $0.482$ & $-0.323$ & $0.486$ & $-0.299$ & $0.492$  & $-0.265$ \\[1ex]
      \hdashline
    \\[0.1ex]
     $+10$  & $0.447$ & $-0.476$ & $0.453$ & $-0.440$ & $0.461$  & $-0.390$ \\[1ex]
    \hline
    \end{tabular}
\caption{The growth index parameters for the different values of $\Omega_{\m0}$ when $M_0 = 0.1$, $\omega_{0} = -1.0$ and $\omega_{a} = 0$ with measurement errors in the growth rate $d \ln \delta / d \ln a$ being $-10$, -5, 5, 10 \%, respectively.} 
\label{tab-3} 
\end{table}

\subsection{Results}

Now, one can reconstruct the $f(R)$ gravity models from the obtained values of ($\omega_{0}, \omega_{a}$) and ($\gamma_{0}, \gamma_{a}$) without and with considering the measurement errors on $\gamma$. One can numerically solve the equation (\ref{FpoF0}) by using the initial condition $F/F_{0} = 1$ and values of $\omega_{0}, \omega_{a}, \gamma_{0}$, and $\gamma_{a}$ given in the tables \ref{tab-2} and \ref{tab-3}. 

\begin{enumerate}[(I)]

\item Without measurement error  : \\

\begin{figure*}
\centering
 \begin{tabular}{cc}
    \includegraphics[width=0.5\linewidth]{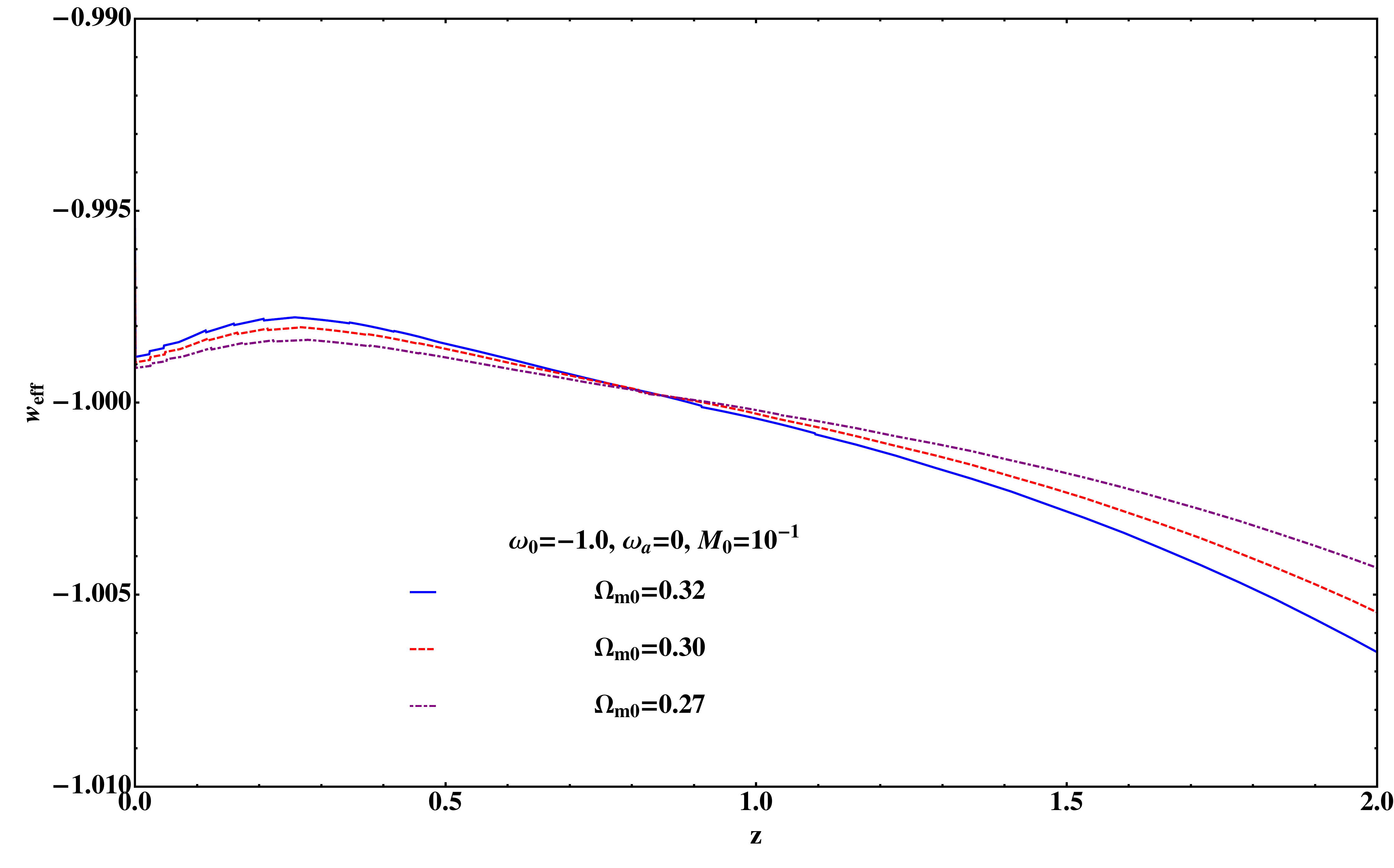} &
    \includegraphics[width=0.5\linewidth]{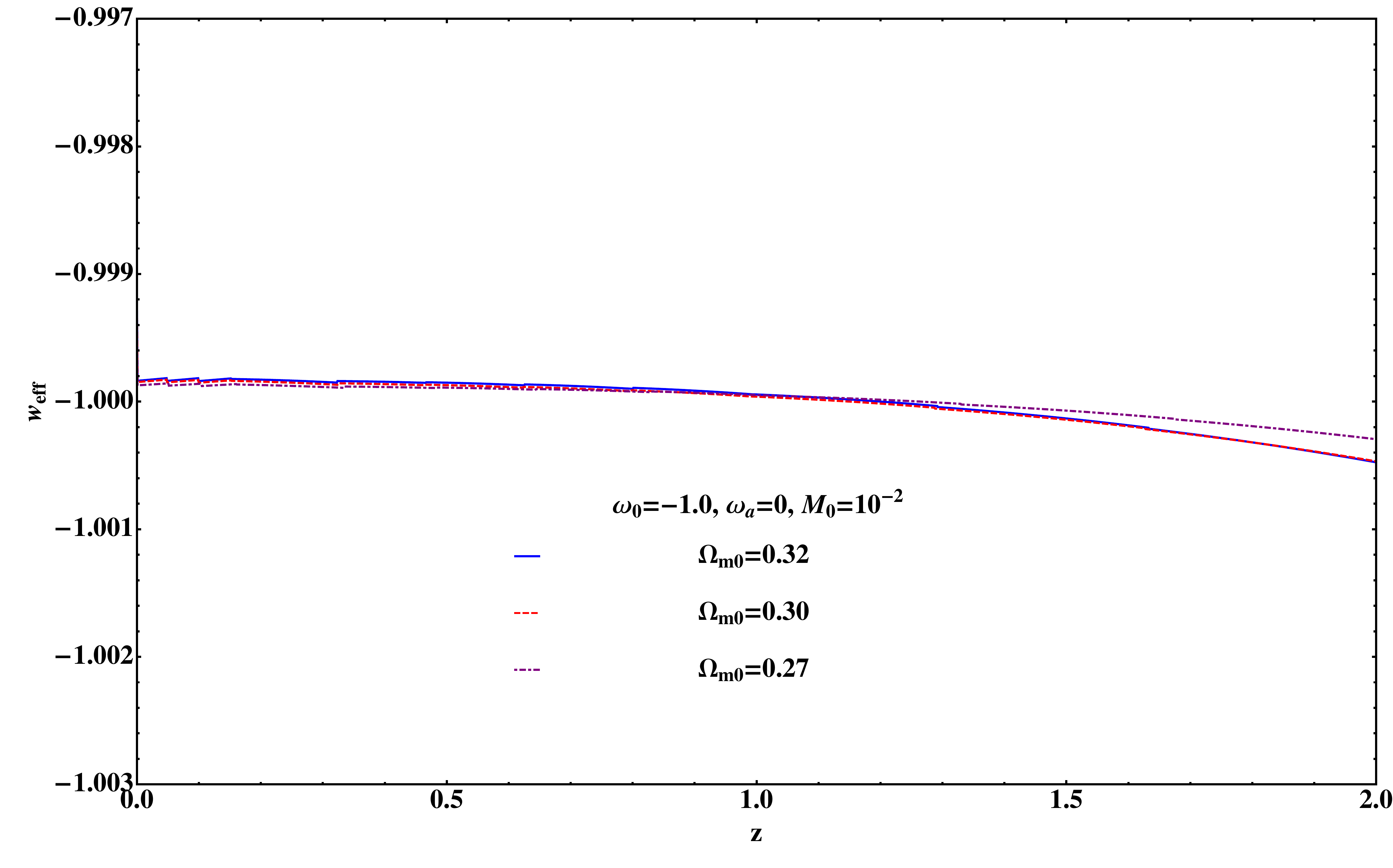}
 \end{tabular}
\caption{Obtained effective DE e.o.s for different models when ($\omega_{0}, \omega_{a}$) = (-1.0, 0). {\it Left panel} : This is for $M_0 = 0.1$. The solid, dashed, and dot-dashed lines correspond $\Omega_{\m0} = 0.32, 0.30$, and 0.27, respectively. {\it Right panel} : $M_0 = 10^{-2}$. We use the same notations as those of left ones for the different models.}
\label{fig-1}
\end{figure*}

First, we investigate the model reconstruction without considering measurement error. We probe the effective DE e.o.s, $\omega_{\eff}$ given in Eq. (\ref{omegaDEMG}). If we assume that all the necessary measurements are measured without errors, then one can reconstruct $\omega_{\eff}$ without any uncertainty. In Fig.\ref{fig-1},  we show the evolution of the effective DE e.o.s for the different values of $M_{0}$ and $\Omega_{\m0}$ when the background evolution mimics that of $\Lambda$CDM. In the left panel of Fig.\ref{fig-1}, we adopt ($\omega_{0}, \omega_{a}, M_{0}$) = (-1.0, 0, 0.1). For different values of $\Omega_{\m0} = (0.32, 0.30, 0.27)$, one obtains ($\gamma_{0}, \gamma_{a}$) = (0.519, -0.162), (0.522, -0.151), (0.525, -0.135) as shown in the table \ref{tab-2}. Thus, one can solve the differential equation (\ref{FpoF0}) to obtain evolution behaviors of $\omega_{\eff}$ given in Eq. (\ref{omegaDEMG}). The solid, dashed, and dot-dashed lines depict $\omega_{\eff}$ for $\Omega_{\m0} = 0.32, 0.30$, and 0.27, respectively. All the models consistent with $\omega_{\eff} = -1.0$ within percent level accuracies. Even though the larger the value of $\Omega_{\m0}$, the larger the deviation of $\omega_{\eff}$ from $\omega = -1.0$, all of those deviations are sub-percent level and one will not be able to distinguish $\omega_{\eff}$ from -1. In the right panel of Fig.\ref{fig-1}, we show the effective DE e.o.s for $M_{0} = 10^{-2}$ with the same notations as the left panel of that figure. One obtains the smaller deviation of $\omega_{\eff}$ from -1 with smaller values of $M_{0}$. This fact is shown in the right panel of Fig.\ref{fig-1}. 

\begin{figure*}
\centering
 \begin{tabular}{cc}
    \includegraphics[width=0.5\linewidth]{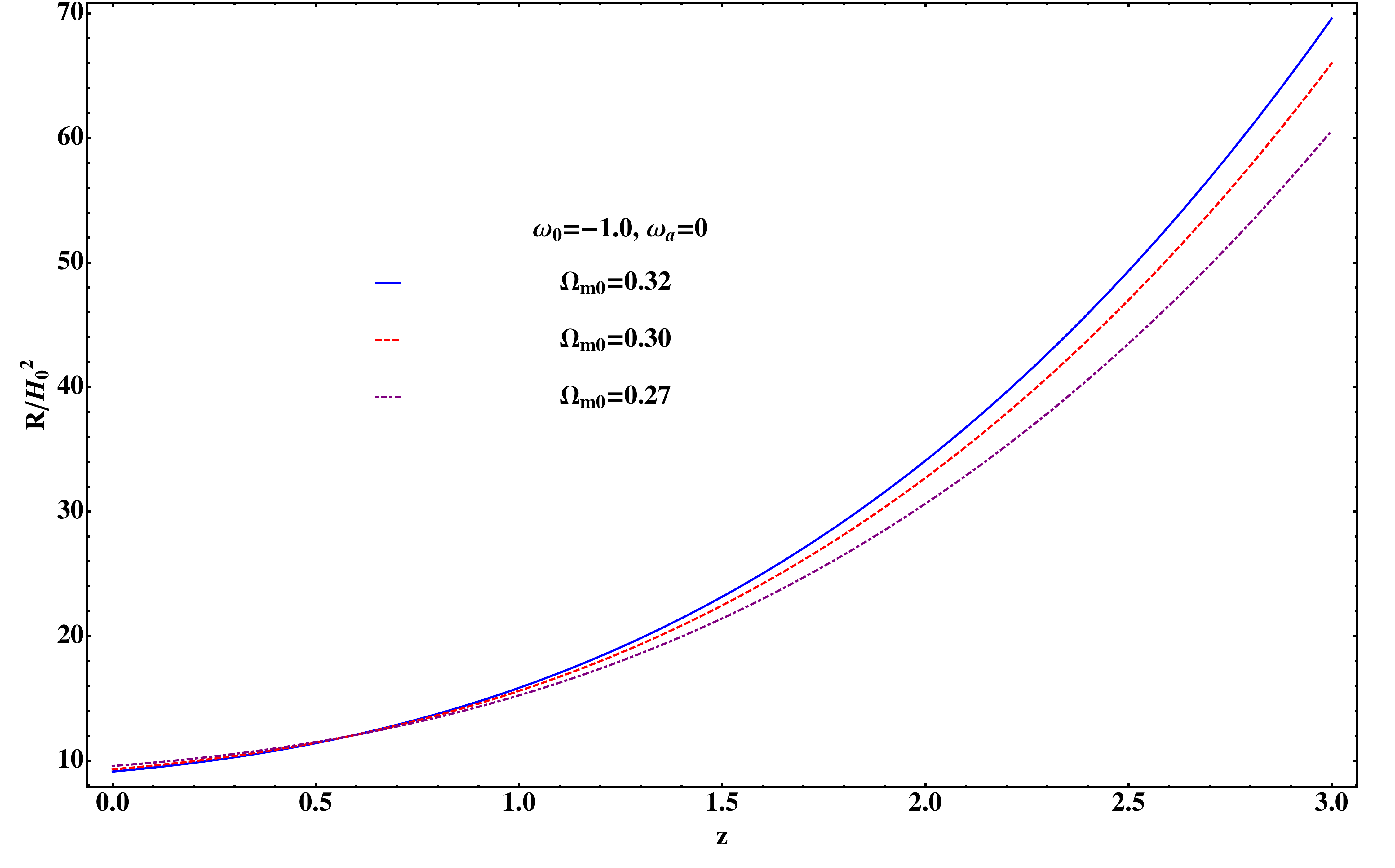} &
    \includegraphics[width=0.5\linewidth]{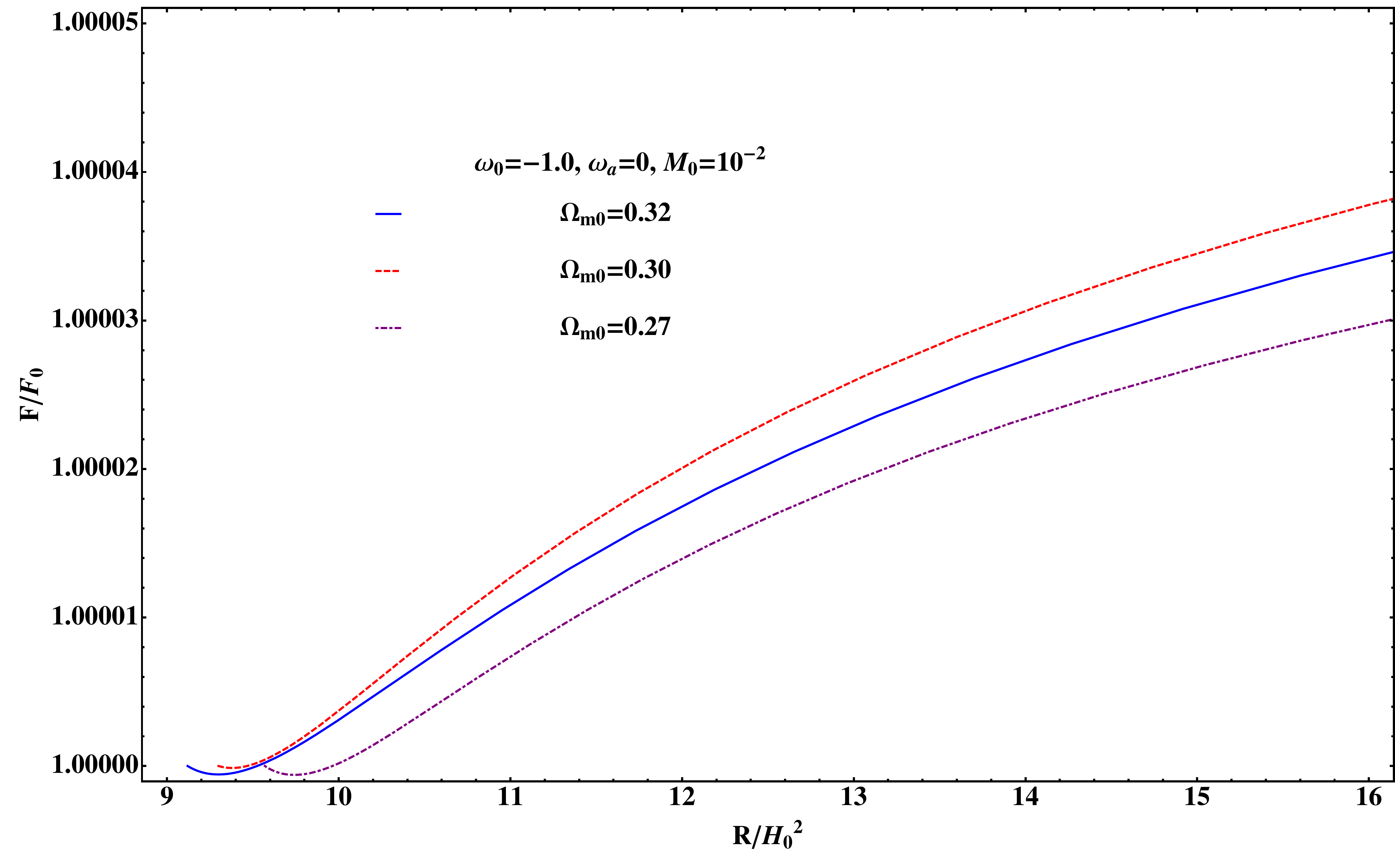} \\
    \includegraphics[width=0.5\linewidth]{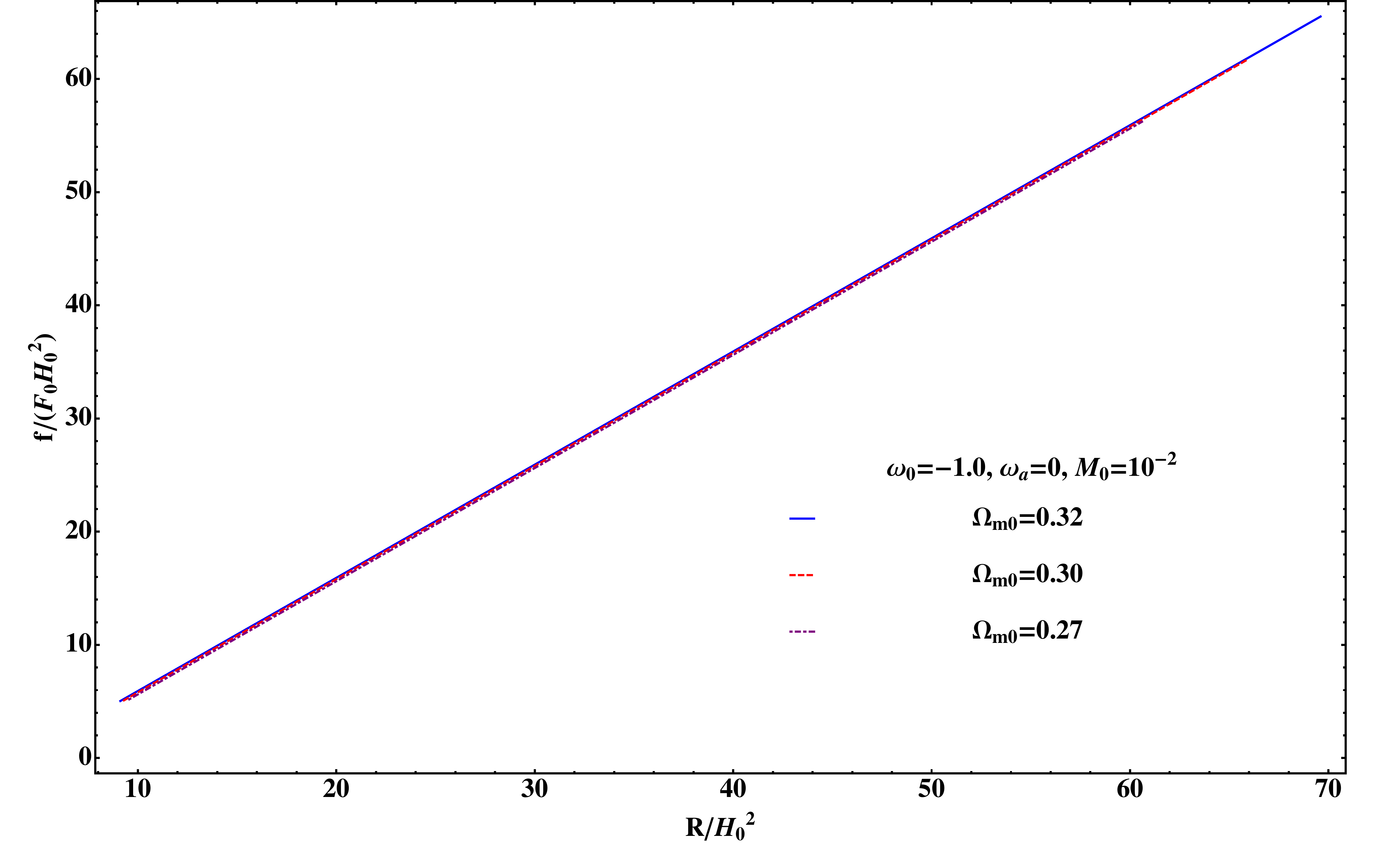} &
    \includegraphics[width=0.5\linewidth]{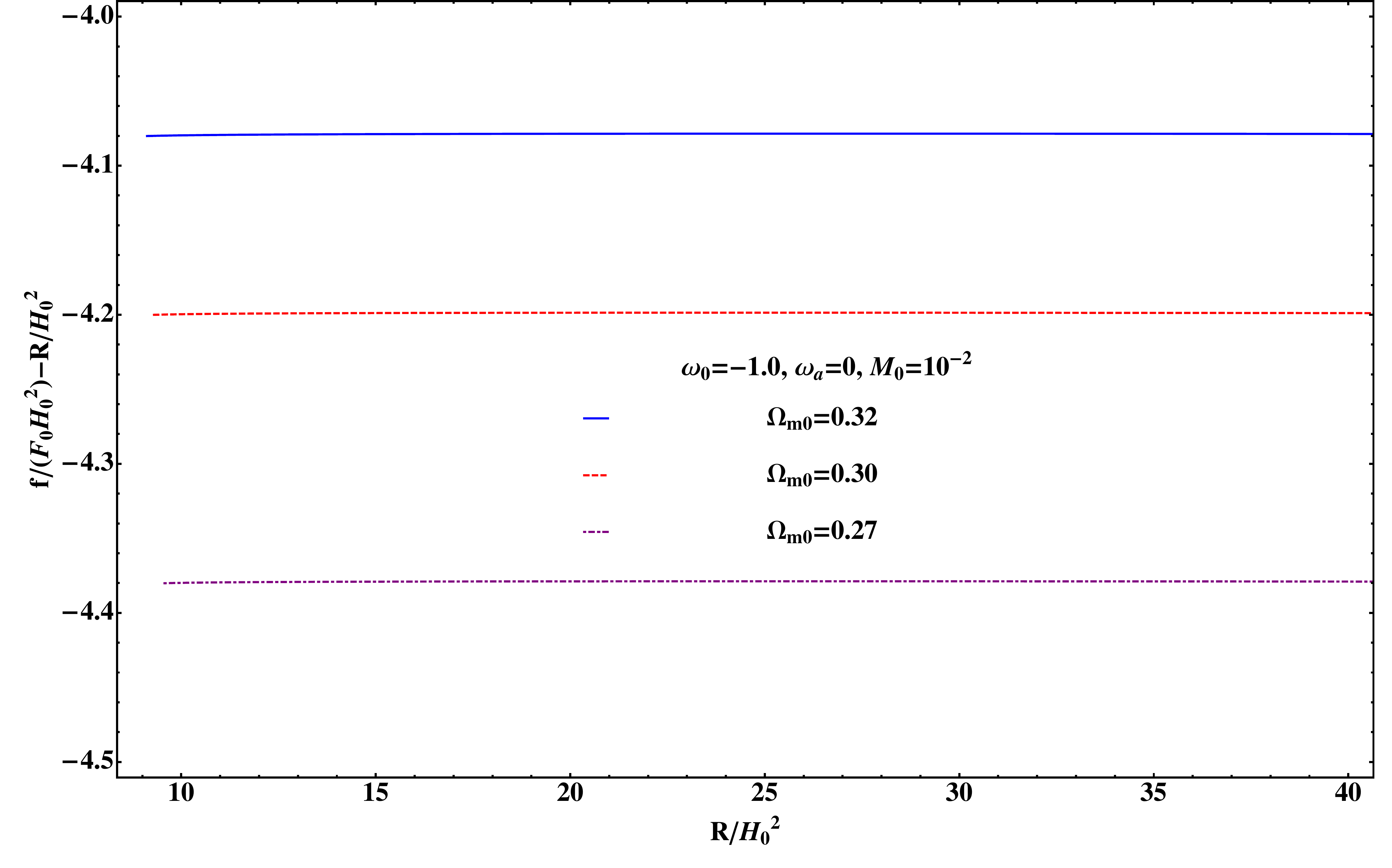}
 \end{tabular}
\caption{Evolutions of several quantities. We adopt the same notations for models. The solid, dashed, and dot-dashed lines correspond $\Omega_{\m0} = 0.32, 0.30$, and 0.27, respectively. {\it Upper left panel} : The evolutions of $R/H_0^2$ for different models as a function of $z$. {\it Upper right panel} : The evolutions of $F/F_0$ for the different values of $\Omega_{\m0}$ as a function of $R/H_0^2$ when $M_0 = 10^{-2}$. {\it Lower left panel} : The evolutions of $f/(F_0 H_0^2$ as a function of $R/H_0^2$. {\it Lower right panel} : The evolutions of $f/(F_0 H_0^2 - R/H_0^2$ as a function of $R/H_0^2$. }
\label{fig-2}
\end{figure*}    

Now, one can reconstruct model functions as a function of the Ricci scalar. In other words, we show the behaviors of $f/(F_0 H_{0}^2)$ and $F/F_{0}$ as a function of $R/H_{0}^2$. For this purpose, we first show the behaviors of $R/H_{0}^2$ as a function of redshift, $z$ for the different values of $\Omega_{\m0}$. This is given by Eq.(\ref{RoH02}). We depict the dimensionless normalized Ricci scalar, $R/H_0^2$ as a function of $z$ in the upper left panel of Fig.\ref{fig-2}. For $z \leq 0.58$, the smaller the values of $\Omega_{\m0}$, the larger the value of $R/H_0^2$. For larger values of $z$, $R/H_{0}^2$ are larger for the larger values of $\Omega_{\m0}$. The solid, dashed, and dot-dashed lines correspond $\Omega_{\m0} = 0.32, 0.30$, and 0.27, respectively. Thus, for the same redshift period, the larger value of $\Omega_{\m0}$ covers the wider range of $R/H_{0}^2$. $R/H_{0}^2$ varies from 9.12 (9.3, 9.8) to 15.8 (15.6, 15.2) for $\Omega_{\m0} = 0.32$ (0.30, 0.27) during redshift $0 \leq z \leq 1.0$. The evolutions of $F/F_{0}$ for the different values of $\Omega_{\m0}$ as a function of $R/H_{0}^2$ are shown in the upper right panel of Fig.\ref{fig-2}. As the initial condition $F/F_{0} |_{n=0} = 1$ is required. We put $M_{0} = 10^{-2}$ in this case. Deviation of $F/F_{0} $ from 1 indicates the modification of the Hilbert-Einstein action. To be consistent with current observation, the deviation of $ \partial f(R)/ \partial R$ from 1 is an order of magnitude $-5$ in these models as shown in the figure. The solid (dashed, dot-dashed) line corresponds to $\Omega_{\m0} = 0.32$ (0.30, 0.27). $F/F_{0}$ varies most rapidly for $\Omega_{\m0} = 0.30$ model. We also shows the behaviors of dimensionless normalized $f/(F_{0} H_{0}^2)$ in the lower left panel of Fig.\ref{fig-2}. With $M_{0} = 10^{-2}$, the solid, dashed, and dot-dashed lines correspond $\Omega_{\m0} = 0.32, 0.30$, and 0.27, respectively. The larger the value of $\Omega_{\m0}$, the larger the magnitude of $f/(F_{0}H_{0}^2)$. The deviations of action from the Hilbert-Einstein action for the different models are depicted in the lower right panel of Fig.\ref{fig-2}. This is given by
\be \fr{f}{F_{0} H_{0}^2} - \fr{R}{H_0^2} = \fr{R}{H_0^2} \left( \fr{f}{F_{0} R} - 1 \right) \label{tf} \, . \ee
Thus, the values shown in the figure show the magnitude of the deviation of $f$ from $R$ multiplied by the normalized Ricci scalar. As shown in the figure, the observations show that the deviation should be negative and the variation of it must be very small. The variation is around -4.1 (-4.2, -4.4) for $\Omega_{\m0} = 0.32$ (0.30, 0.27).   

\item With measurement error : \\

\begin{figure*}
\centering
 \begin{tabular}{cc}
    \includegraphics[width=0.5\linewidth]{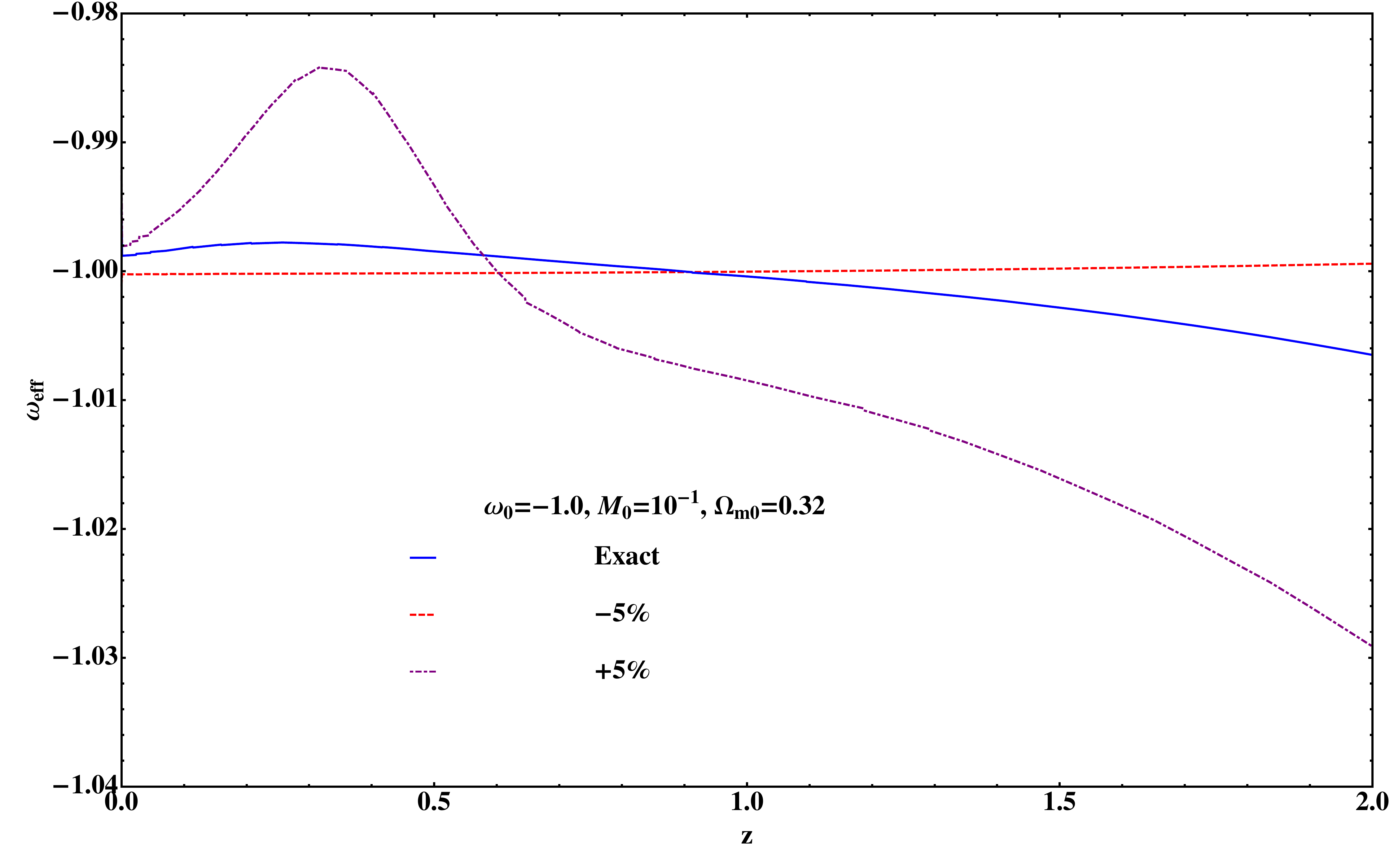} &
    \includegraphics[width=0.5\linewidth]{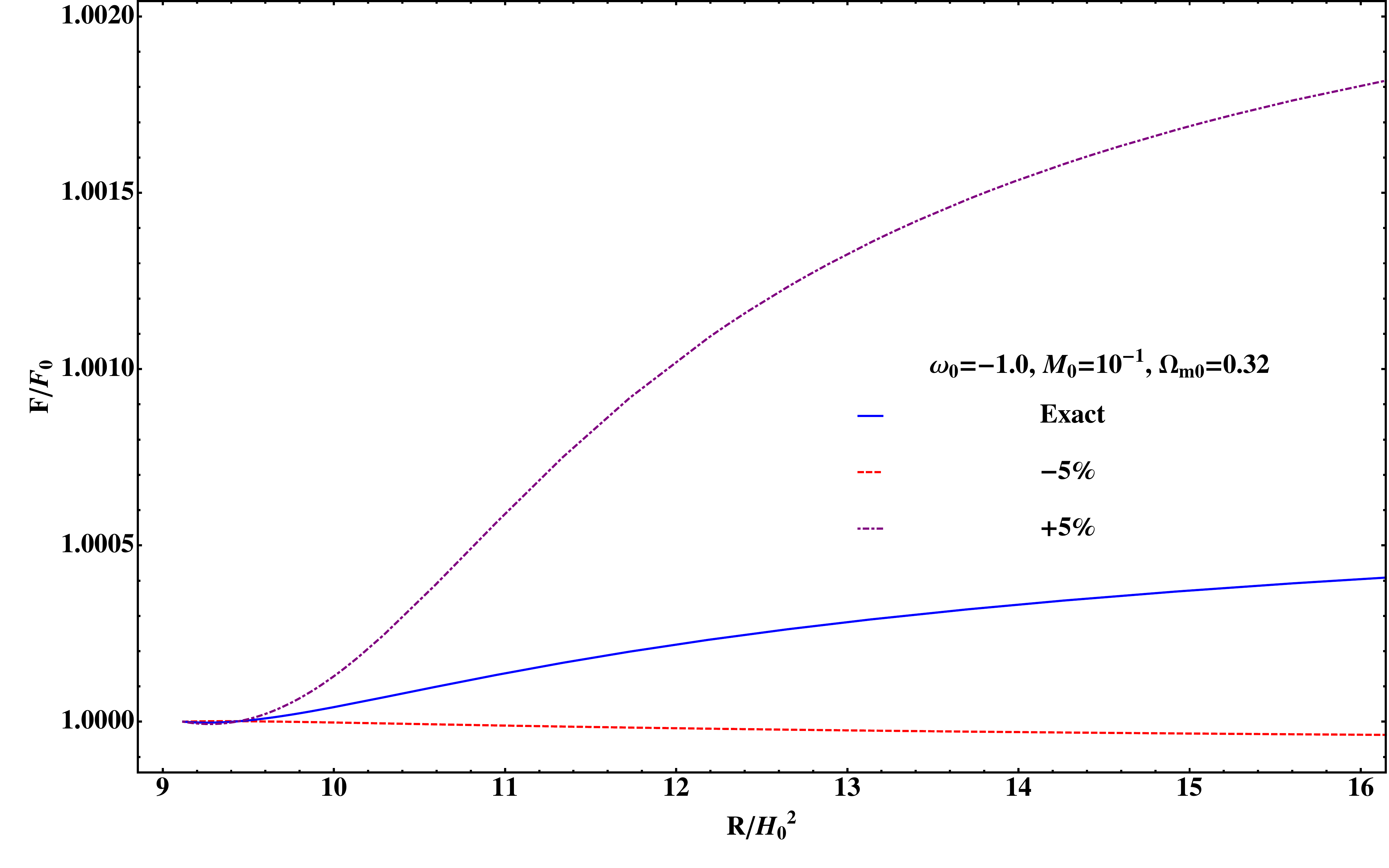} 
 \end{tabular}
\caption{Evolutions of $\omega$ and $F/F_{0}$ including measurement errors on the growth rate when $(\omega_{0}, \Omega_{\m0}, M_{0}) = (-1.0, 0.32, 0.1)$. {\it Left panel} : The evolutions of $\omega_{\eff}$ if we include the measurement errors on $d \ln \delta / d \ln a$. The dashed line and the dot-dashed line correspond the measurement errors on the growth rate -5 \% and + 5 \%, respectively. {\it Right panel} : The evolutions of $F/F_{0}$ as a function of $R/H_{0}^2$ with the measurement errors on the growth rate. The dashed (dot-dashed) line correspond -5 (+5) \% measurement errors on $d \ln \delta / d \ln a$.}
\label{fig-3}
\end{figure*}  

In the previous consideration, we do not include measurement errors in both $\omega$ and $\gamma$. However, in the real observations, measured values of them include measurement errors. Compared to $\gamma$, one might ignore the errors on $\omega$. In future observations, the error on the growth rate $d \ln \delta / d \ln a$ might be reduced to 5 - 10 \% level. Thus, we consider the $f(R)$ model reconstruction including measurement errors on the growth rate. We numerically obtain the growth index parameters with including measurement errors on $d \ln \delta / d \ln a$ in the table \ref{tab-3}. We obtain ($\gamma_{0}, \gamma_{a}$) for the different models with the measurement errors on the growth rate from -10 \% to + 10 \%. For example, when the measurement errors on the growth rate is +5 \%, then ($\gamma_{0}, \gamma_{a}$) = (0.482, -0.323) when $\Omega_{m0} = 0.32$.  As shown in the table \ref{tab-3}, if the measurements on the growth rate become smaller than the true values, then both $\gamma_{0}$ and $\gamma_{a}$ become larger than true values for all models. Also if the measurements on the growth rate are larger than the true values, then both $\gamma_{0}$ and $\gamma_{a}$ are smaller than true values of them.

We show the evolutions of $\omega_{\eff}$ and $F/F_{0}$ with including the measurement errors on the growth rate in Fig.\ref{fig-3}. We investigate the models with $(\omega_{0}, \Omega_{\m0}, M) = (-1.0, 0.32, 0.1)$. In the left panel of Fig.\ref{fig-3}, we show the evolutions of the effective DE e.o.s when we include the measurement errors on $d \ln \delta / d \ln a$. The dashed line correspond the case when the measurement of the growth rate is 5 \% smaller than the true value. The smaller values of the growth rate than the true one induce the smaller change in the modification terms of the model and thus produce smaller change in $\omega_{\eff}$. The dot-dashed line represents the case when the growth rate measurement is 5\% larger than that of the true value. In this case, the larger deviation of the $F/F_{0}$ induces the larger deviation of $\omega_{\eff}$ from -1. The deviation can be as large as 3 \% at $z \sim 2$ in this case. Thus, one might have inconsistent values of $\omega_{\eff}$ obtained from the background evolution and the large scale structure. In the right panel of Fig.\ref{fig-3}, we show the evolutions of $F/F_0$ when we include the measurement errors. The dot-dashed line indicates the evolution of $F/F_0$ when the measurement error on the growth rate is +5 \%. This case shows the rapid change of $F/F_0$ compared to the real model. This might give the totally different functional form of the model compared to the original model. Also when the error on the growth rate is - 5\%, the $F/F_0$ can be smaller than 1. And this case, one might conclude the gravitational force is smaller than the present one. This gives the opposite result compared to the original model. Thus, the interpretation of $f(R)$ model with measurement errors requires more carel.

\end{enumerate}

\subsection{Comparison with Viable model} 
\begin{figure*}
\centering
    \includegraphics[width=0.7\linewidth]{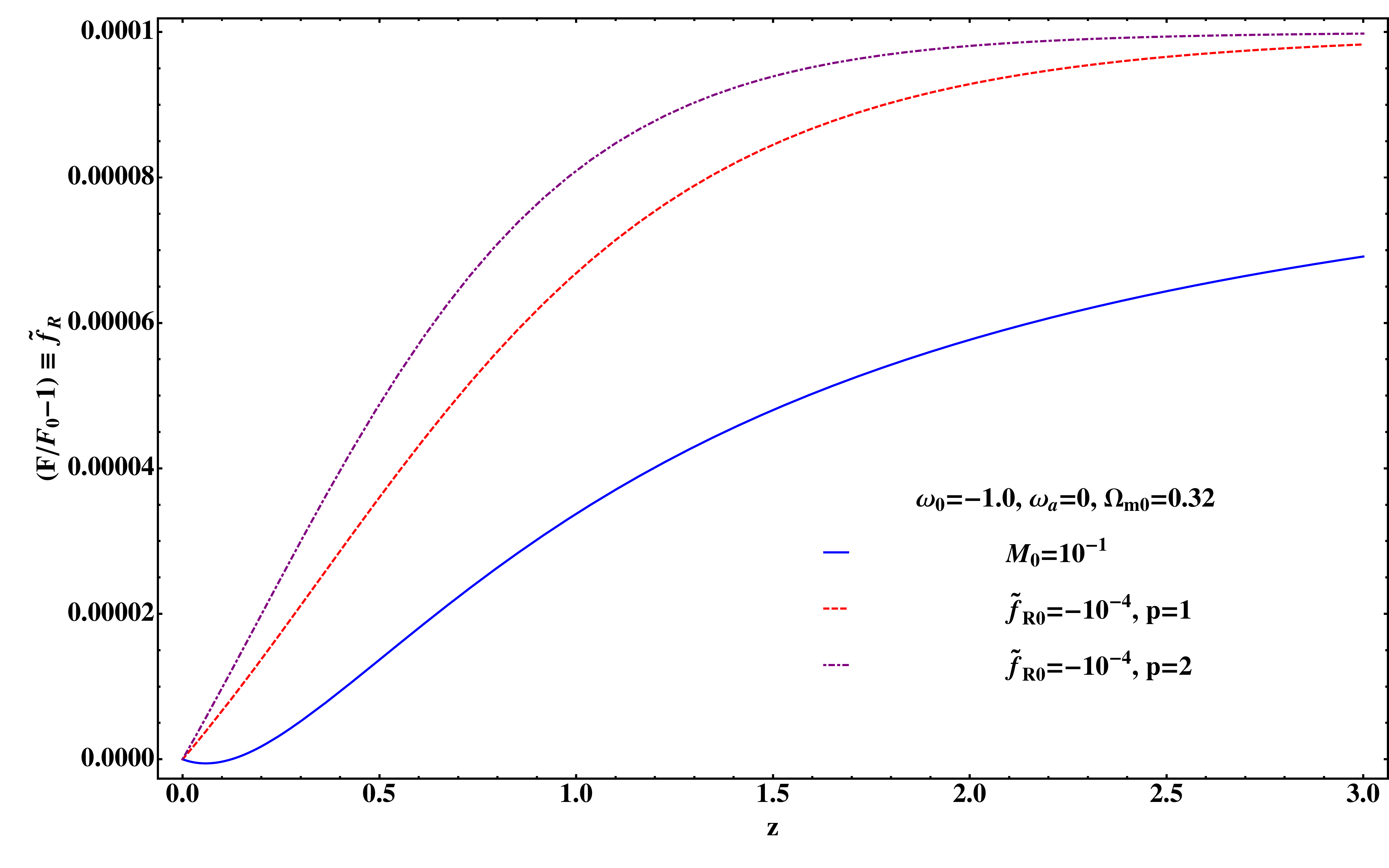} 
\caption{Comparison of our reconstruction method with HS model when $(\omega_{0}, \omega_{a}, \Omega_{\m0}) = (-1, 0, 0.32)$ . The solid line depicts the evolution of reconstructed $\tilde{f}_{R}$ when $M_{0} = 10^{-2}$.  The dashed (dot-dashed) line shows the evolution of it when $p = 1$ (2) with $\tilde{f}_{R0} = 10^{-4}$.}
\label{fig-4}
\end{figure*}  
We list the viable $f(R)$ gravity models in the table\ref{tab-1}. We compare the reconstructed models with the so-called ``Hu-Sawicki (HS)'' model \cite{2007PhRvD..76f4004H}. If one converts the notations of HS model to ours, then one obtains
\ba R_{\HS} &=& F_{0} H_{0}^2 \Omega_{\m0} \, , \label{RHS} \\
\fr{c_{1}}{c_{2}} &=& 6 \fr{1-\Omega_{\m0}}{\Omega_{\m0}} \, , \label{c1} \\
c_{2} &=& -p \fr{c_{1}}{c_{2}} \left( \fr{12}{\Omega_{\m0}} -1 \right)^{-p-1} \fr{1}{\tilde{f}_{R0}} \label{c2} \, , \ea
where $p$ is the natural number and $\tilde{f}_{R0}$ is the initial input parameter. The figure \ref{fig-4} depicts the evolutions of $\tilde{f}_{R} = F/F_0 -1$ for both the reconstructed model and the HS models. We adopt $(\omega_{0}, \omega_{a}, \Omega_{\m0}) = (-1, 0, 0.32)$ in this figure. The solid line indicates the evolution of reconstructed $\tilde{f}_{R}$ when $M_{0} = 10^{-2}$.  The dashed (dot-dashed) line shows the evolution of $\tilde{f}_{R}$ of HS models when $p = 1$ (2) with $\tilde{f}_{R0} = 10^{-4}$. All of them show the same behaviors as $F/F_0 -1$ increase as $z$ increases. The main difference between our reconstruction method and HS model is the rate of $\tilde{f}_{R}$ which depends on $\gamma_{0}$ and $\gamma_{a}$ in our method. The reconstruction method can provide general function of $f(R)$ gravity models which consistent with observations.  

\section*{Conclusions}
We illustrate how to determine the general $f(R)$ gravity models from cosmological observations. Thus, if cosmological observables are accurate enough, then one can constrain the model functions. Especially, if the growth rate index is determined with high accuracy, then one can well specify the model. Even if the future observations are consistent with $\Lambda$CDM in background evolutions, one might have a chance to specify $f(R)$ gravity if the growth rate of the large scale structure is different from that of $\Lambda$CDM. Our method depends on only observations and can be used for various model search after the more accurate large scale structure data are available in the near future. 


\appendix
\section{}
\setcounter{equation}{0}
One can rewrite evolution equations for both the background and the matter perturbation Eqs.(\ref{G00})-(\ref{dotdeltam}) as a function of $n \equiv \ln a$ in order to relate the$f(R)$ gravity models as a function of observed quantities. 

\subsection{Background Evolution}
First, we rewrite the background evolution equations Eqs.(\ref{G00})-(\ref{Gii}) as a function of $n$  
\ba 3 F_{0} H^2 &=& \fr{1}{2} \left( FR - f \right) - 3H^2 \left(F' + F - F_{0} \right) + 8 \pi G \rho_{\m} \, , \label{G00n} \\
 -2 F_{0} H H' &=& H^2 \left( F'' - F' \right) + H H' \left(F' +2F - 2F_{0} \right) + 8 \pi G \rho_{\m} \, , \label{Giin} \ea
where primes mean the derivative with respect to $n$. One can rewrite the above equations 
\ba 3 \fr{H^2}{H_{0}^2} &=& \fr{1}{2} \left( \fr{F}{F_0} \fr{R}{H_0^2} - \fr{f}{F_0 H_0^2} \right) - 3\fr{H^2}{H_0^2} \left(\fr{F'}{F_0} + \fr{F}{F_0} - 1 - \Omega_{\m} \right)  \, , \label{G00n2A} \\
 -2 \fr{H'}{H} &=& \left( \fr{F''}{F_0} - \fr{F'}{F_0} \right) + \fr{H'}{H} \left(\fr{F'}{F_0} +2\fr{F}{F_0} - 2 \right) + 3 \Omega_{\m} \, , \label{Giin2A} \ea
From the above Eqs.(\ref{G00n}) and (\ref{Giin}), one can define the energy density, $\rho_{\eff}$ the pressure, $P_{\eff}$, and the equation of state, $\omega_{\eff}$ of the effective dark energy derived from the modification of the theory
\ba \rho_{\eff} &=& \fr{H^2}{8\pi G} \left( \fr{\left( FR - f \right)}{2H^2}  - 3\left(F' + F - F_{0} \right) \right)\, , \label{rhoeff} \\
P_{\eff} &=& \fr{H^2}{8\pi G} \left( \fr{\left( f-FR \right)}{2H^2} + \left(F'' + 2F' +3F -3F_{0}\right) \right) \nonumber \\ 
&+& \fr{H^2}{8\pi G} \left( \fr{H'}{H} \left(F'+2F -2F_{0}\right) \right) \, , \label{Peff} \\ 
\omega_{\eff} &=& -1 + \fr{H^2 \left( F'' - F' \right) + H H' \left(F' +2F - 2F_{0} \right) }{\fr{1}{2} \left( FR - f \right) - 3H^2 \left(F' + F - F_{0} \right)} \nonumber \\ 
&=& -1 + \fr{H^2 \left( F'' - F' \right) + H H' \left(F' +2F - 2F_{0} \right) }{3F_{0} H^2 \left( 1 - \Omega_{\m} \right)} \, . \label{weff} \ea
We adopt the so-called Chevallier-Polarski-Linder (CPL) parameterization of the dark energy e.o.s $\omega_{\DE}$ \cite{2001IJMPD..10..213C, 2003PhRvL..90i1301L} to match the background evolution equations (\ref{G00n})-(\ref{Giin})
\ba 3 F_{0} H^2 &=& 8\pi G \left( \rho_{\m} + \rho_{\DE} \right) \equiv 8\pi G \rho_{\crt} \label{G00cpl} \, , \\
-2 F_{0} H H' &=& 8\pi G \left( \rho_{\m} + \rho_{\DE} + P_{\DE} \right) \label{Giicpl} \, , \\
\rho_{\DE} &=& \rho_{\DE 0} e^{-3\left(1+\omega_0+\omega_a \right)n + 3\omega_{a}\left(e^{n}-1\right)} \label{rhocpl} \, , \\ 
P_{\DE} &=& \omega_{\DE} \rho_{\DE} \label{Pcpl} \, , \ea
where $\rho_{\crt}$ is the critical density. Thus, one can parameterize model $f(R)$ and its derivatives with respect to $n$ as a function of $\omega_{0}$ and $\omega_{a}$ 
\ba \Omega_{\DE}\left[\Omega_{\m0}, \omega_{0}, \omega_{a}, n \right] &=& \fr{1}{6} \left( \fr{F}{F_0} \fr{R}{H_{0}^2} - \fr{f}{F_{0} H_{0}^2} \right) \fr{H_{0}^{2}}{H^2} - \left(\fr{F'}{F_0} + \fr{F}{F_0} - 1 \right)  \label{OmgeaDEMG} \, , \\
&\equiv& \fr{g\left[\Omega_{\m0}, \omega_{0}, \omega_{a}, n \right] }{1+g\left[\Omega_{\m0}, \omega_{0}, \omega_{a}, n \right] } \equiv 1 - \Omega_{\m} \, , \nonumber \\
g\left[\Omega_{\m0}, \omega_{0}, \omega_{a}, n \right] &=& \fr{1-\Omega_{\m0}}{\Omega_{\m0}} e^{-3\left(\omega_{0}+\omega_{a}\right)n +3\omega_{a}\left(e^n-1\right)} = \fr{\Omega_{\DE}}{\Omega_{\m}} \, , \label{gwowa} \\
1 + \omega\left[\omega_{0},\omega_{a},n\right] &=& \fr{\left( \fr{F''}{F_0} - \fr{F'}{F_0} \right) + \fr{H'}{H} \left(\fr{F'}{F_0} +2\fr{F}{F_0} -2\right) }{3\Omega_{\DE}\left[\Omega_{\m0}, \omega_{0}, \omega_{a},n \right]} \label{omegaDEMG} \, . \ea 
Thus, one can also rewrite the Friedmann equations 
\ba \fr{H^2}{H_0^2} &=& \fr{\rho_{\m}}{\rho_{\crt 0}} \left( 1 + \fr{\rho_{\DE}}{\rho_{\m}} \right) = \Omega_{\m 0} e^{-3n} \left(1+g\left[\Omega_{\m0}, \omega_{0}, \omega_{a},n \right] \right) \label{H2oH02A} \, , \\
\fr{H'}{H} &=& -\fr{3}{2} \left( 1 + \omega \Omega_{\DE} \right) \equiv -\fr{3}{2} \left(1 + Q\left[\Omega_{\m0}, \omega_{0}, \omega_{a}, n \right]  \right)\label{HoHpA} \, , \\
\fr{R}{H_{0}^2} &=& 6 \fr{H^2}{H_0^2} \left( 2 + \fr{H'}{H} \right) = 3 \Omega_{\m0} e^{-3n} \left(1+g\right) \left(1-3Q\right) \label{RoH02} \, . \ea 
Even though one represents $f$, $F =\equiv \fr{\partial f}{\partial R}$, and derivatives of $F$ with respect to $n$ as a function of $\omega_{\DE}$, this is not enough to constrain the evolution equations of them. Thus, one needs to use further constraints from the perturbation evolution equations to specify the models.

\subsection{Perturbation Evolution}
We also rewrite the matter perturbation equation (\ref{deltam}) as a function of $n$
\be
 \delta_{\m}^{''} = - \left( 2 + \fr{H'}{H} \right) \delta_{\m}' + \fr{3}{2} \fr{F_{0}}{F} \Omega_{\m} \left( \fr{1+4M}{1+3M} \right) \delta_{\m} \, . \label{deltamnA} \ee
One can parametrize the growth rate of the matter perturbation, $d \ln \delta_{\m} / d \ln a \equiv \Omega_{\m}^{\gamma}$ and the growth rate index, $\gamma$ can be parameterized as \cite{2011JCAP...03..021L}
\be \gamma = \gamma_{0} + \gamma_{a} \left(1 - e^{n} \right) \, . \label{gamma} \ee 
By using Eqs.(\ref{H2oH02})-(\ref{gamma}), one rewrite the Eq.(\ref{deltamn}) to obtain
\ba {\cal P}\left[\Omega_{\m0}, \omega_{0}, \omega_{a}, \gamma_{0}, \gamma_{a}, n \right]  &\equiv& \Omega_{\m}^{\gamma} \left( \Omega_{\m}^{\gamma} + \gamma' \ln \Omega_{\m} + 3 \gamma Q + \fr{\left(1-3Q\right)}{2} \right) \label{PA} \, , \\
\fr{F\left[\Omega_{\m0}, \omega_{0}, \omega_{a}, \gamma_{0}, \gamma_{a}, k, n \right] }{F\left[\Omega_{\m0}, \omega_{0}, \omega_{a}, \gamma_{0}, \gamma_{a}, k, 0 \right]} &=& \fr{3}{2} \fr{\Omega_{\m}}{{\cal P}} \left(\fr{1+4M}{1+3M}\right) \equiv \fr{F}{F_0} \label{FoF0nA} \, , \\ 
M = \fr{k^2}{a^2 H_{0}^2} \fr{H_0^2}{R'} \fr{F'}{F} &=& \fr{1-A}{3A -4} \,\, \text{where} \,\, A = \fr{2}{3}\fr{F}{F_{0}} \fr{{\cal P}}{\Omega_{\m}} \label{MA} \, , \\
\fr{F'}{F_{0}} = \fr{a^2 H_{0}^2}{k^2} \fr{R'}{H_{0}^2} M \fr{F}{F_0} &=& \fr{a^2 H_{0}^2}{k^2} \fr{R'}{H_{0}^2} \left( \fr{1-A}{3A-4} \right) \fr{F}{F_0} \label{FpoF0A} \, . \ea
One can solve non-linear differential equation (\ref{FpoF0A}) to solve the evolution of $F/F_{0}$ with the initial condition $F[n=0]/F_{0} =1$.
Both $\omega[\omega_0,\omega_a]$ and $\gamma[\gamma_0,\gamma_a]$ can be obtained from cosmological observations and thus one can obtain the k-dependent $F/F_0$ values from the observations as shown in Eq.(\ref{FoF0n}).

The below table shows the all the required quantities as functions of $\omega_{0}, \omega_{a}, \gamma_{0}$, and $\gamma_{a}$. 

\begin{table}
\centering
\begin{tabular}{c cc}
\hline \hline \\
Obtained Quantities & Measured Quantities  & Functions \\[2ex]		
\hline
\\[0.2ex]
$\Omega_{\m}\left[\Omega_{\m0}, \omega_{0}, \omega_{a},n\right]$ & $\Omega_{\m0}, \omega_{0}, \omega_{a}$  &  $\left( 1 + \fr{1-\Omega_{\m0}}{\Omega_{\m0}} e^{-3\left(\omega_0+\omega_a \right)n + 3\omega_{a}\left(e^{n}-1\right)}  \right)^{-1}$ \\[3ex]
\hdashline
\\[0.2ex]
$\fr{H^2}{H_{0}^2} = \fr{H^2\left[\Omega_{\m0}, \omega_{0}, \omega_{a},n\right]}{H^2\left[\Omega_{\m0}, \omega_{0}, \omega_{a},0\right]}$ & $\Omega_{\m0}, \omega_{0}, \omega_{a}$ & $\Omega_{\m0} e^{-3n} + \left( 1 - \Omega_{\m0}\right) e^{-3\left(1+\omega_0+\omega_a \right)n + 3\omega_{a}\left(e^{n}-1\right)}  $ \\[3ex]
\hdashline
\\[0.2ex]
$Q\left[\Omega_{\m0}, \omega_{0}, \omega_{a},n\right]$ & $\Omega_{\m0}, \omega_{0}, \omega_{a}$ & $\omega\left[\omega_{0},\omega_{a},n\right] \left(1-\Omega_{\m}\left[\Omega_{\m0}, \omega_{0}, \omega_{a}, n \right] \right) $ \\[3ex]
\hdashline
\\[0.2ex]
$\fr{H^{'}\left[\Omega_{\m0}, \omega_{0}, \omega_{a},n\right]}{H\left[\Omega_{\m0}, \omega_{0}, \omega_{a},n\right]}$  &  $\Omega_{\m0}, \omega_{0}, \omega_{a}$ & $-\fr{3}{2} \left( 1 + Q\left[\Omega_{\m0}, \omega_{0}, \omega_{a},n\right]\right) $ \\[3ex]
\hdashline
\\[0.2ex]
$\fr{R\left[\Omega_{\m0}, \omega_{0}, \omega_{a},n\right]}{H_{0}^2\left[\Omega_{\m0}, \omega_{0}, \omega_{a},n\right]}$  &  $\Omega_{\m0}, \omega_{0}, \omega_{a}$ & $6 \fr{H^2\left[\Omega_{\m0}, \omega_{0}, \omega_{a},n\right]}{H^2\left[\Omega_{\m0}, \omega_{0}, \omega_{a},0\right]} \left(2 + \fr{H^{'}\left[\Omega_{\m0}, \omega_{0}, \omega_{a},n\right]}{H\left[\Omega_{\m0}, \omega_{0}, \omega_{a},n\right]} \right)$ \\[3ex]
\hdashline
\\[0.2ex]
${\cal P}\left[\Omega_{\m0}, \omega_{0}, \omega_{a}, \gamma_{0}, \gamma_{a}, n \right] $  &  $\Omega_{\m0}, \omega_{0}, \omega_{a}, \gamma_{0}, \gamma_{a}$ & $\Omega_{\m}^{\gamma} \left( \Omega_{\m}^{\gamma} + \gamma' \ln \Omega_{\m} + 3 \gamma Q + \fr{\left(1-3Q\right)}{2} \right) $ \\[3ex]
\hdashline
\\[0.2ex]
$\fr{F}{F_{0}} = \fr{F\left[\Omega_{\m0}, \omega_{0}, \omega_{a}, \gamma_{0}, \gamma_{a}, k, n \right]}{F\left[\Omega_{\m0}, \omega_{0}, \omega_{a}, \gamma_{0}, \gamma_{a}, k, 0 \right]} $  &  $\Omega_{\m0}, \omega_{0}, \omega_{a}, \gamma_{0}, \gamma_{a}, k$ & $\fr{3}{2} \fr{\Omega_{\m}}{{\cal P}} \left(\fr{1+4M}{1+3M}\right) $ \\[3ex]
\hdashline
\\[0.2ex]
$A\left[\Omega_{\m0}, \omega_{0}, \omega_{a}, \gamma_{0}, \gamma_{a}, k, n \right]$  &  $\Omega_{\m0}, \omega_{0}, \omega_{a}, \gamma_{0}, \gamma_{a}, k$ & $\fr{2}{3}\fr{F}{F_{0}} \fr{{\cal P}}{\Omega_{\m}} $ \\[3ex]
\hdashline
\\[0.2ex]
$M\left[\Omega_{\m0}, \omega_{0}, \omega_{a}, \gamma_{0}, \gamma_{a}, k, n \right]$  &  $\Omega_{\m0}, \omega_{0}, \omega_{a}, \gamma_{0}, \gamma_{a}, k$ & $\fr{1-A}{3A-4} $ \\[3ex]
\hdashline
\\[0.2ex]
$\fr{F'}{F_{0}} = \fr{F'\left[\Omega_{\m0}, \omega_{0}, \omega_{a}, \gamma_{0}, \gamma_{a}, k, n \right]}{F\left[\Omega_{\m0}, \omega_{0}, \omega_{a}, \gamma_{0}, \gamma_{a}, k, 0 \right]} $  & $\Omega_{\m0}, \omega_{0}, \omega_{a}, \gamma_{0}, \gamma_{a}, k$  &  $\fr{e^{2n} H_{0}^2}{k^2} \fr{R'}{H_{0}^2} M \fr{F}{F_0}$ \\[3ex]
\hdashline
\\[0.2ex]
$\omega \left[\Omega_{\m0}, \omega_{0}, \omega_{a}, \gamma_{0}, \gamma_{a}, k, n \right]$ & $\Omega_{\m0}, \omega_{0}, \omega_{a}, \gamma_{0}, \gamma_{a}, k$  & $-1 + \fr{\left( \fr{F''}{F_0} - \fr{F'}{F_0} \right) + \fr{H'}{H} \left(\fr{F'}{F_0} +2\fr{F}{F_0} -2\right) }{3 \left(1 - \Omega_{\m} \right)}$ \\[3ex]
\hline
\end{tabular}
\caption{Reconstructed model functions as functions of measured quantities, ($\omega_{0}, \omega_{a}, \gamma_{0}, \gamma_{a}$).} 
\label{tab-4}
\end{table}

\section*{Acknowledgments}
SL is supported by Basic Science Research Program through the National Research Foundation of Korea (NRF)
funded by the Ministry of Science, ICT and Future Planning (Grant No. NRF-2015R1A2A2A01004532) and (NRF-2017R1A2B4011168.).


\end{document}